\documentclass[conference,a4paper]{APSIPA2025}
\usepackage{amsmath}
\usepackage{graphicx}
\usepackage{multirow}

\usepackage[utf8]{inputenc}
\usepackage[T1]{fontenc}
\usepackage{microtype}
\usepackage{times}
\usepackage{latexsym}
\usepackage{inconsolata}

\usepackage{amsmath}
\usepackage{amssymb}

\usepackage[table]{xcolor}   
\usepackage{colortbl}        
\usepackage{pgf}             
\usepackage{booktabs}        
\usepackage{multirow}
\usepackage{adjustbox}

\usepackage{verbatim}
\usepackage{subcaption}
\usepackage{algorithm2e}
\usepackage{url}
\usepackage{float}
\usepackage{etoolbox}
\usepackage{tcolorbox}

\definecolor{lightred}{rgb}{1.0, 0.8, 0.8}
\definecolor{lightblue}{rgb}{0.8, 0.9, 1.0}
\definecolor{lightgreen}{rgb}{0.8, 1.0, 0.8}
\definecolor{lightyellow}{rgb}{1.0, 1.0, 0.8}
\definecolor{lightpurple}{rgb}{0.9, 0.8, 1.0}
\definecolor{lightorange}{rgb}{1.0, 0.9, 0.8}
\usepackage{inconsolata}
\usepackage{adjustbox}
\usepackage{ subcaption, algorithm2e, url, float, etoolbox, adjustbox, pgf,booktabs, verbatim}
\usepackage{xcolor}
\usepackage{amsmath}
\usepackage{amssymb}
\usepackage{tcolorbox}

\usepackage[backend=biber,style=ieee,]{biblatex}
\usepackage{geometry}
\usepackage{fancyhdr}

\fancypagestyle{firststyle}{
  \fancyhf{}
  \fancyhead[C]{2025 Asia Pacific Signal and Information Processing Association Annual Summit and Conference (APSIPA ASC)}
}

\begin{document}

\title{Investigating Polyglot Speech Foundation Models for Learning Collective Emotion from Crowds}

\author{
\authorblockN{
Orchid Chetia Phukan\authorrefmark{2}\authorrefmark{1},
Girish\authorrefmark{2}\authorrefmark{3}\authorrefmark{1},
Mohd Mujtaba Akhtar\authorrefmark{2}\authorrefmark{4}\authorrefmark{1},
Panchal Nayak\authorrefmark{5},
Priyabrata Mallick\authorrefmark{6}\\
Swarup Ranjan Behera\authorrefmark{6},
Parabattina Bhagath\authorrefmark{7},
Pailla Balakrishna Reddy\authorrefmark{8},
Arun Balaji Buduru\authorrefmark{2}
}

\authorblockA{
\authorrefmark{2}IIIT-Delhi, India,
\authorrefmark{3}UPES, India,
\authorrefmark{4}V.B.S.P.U, India,
\authorrefmark{5}VIT, India,
\authorrefmark{6}Independent Researcher, India \\
\authorrefmark{7}L.B.R College of Engineering, India,
\authorrefmark{8}Reliance AI, India\\
E-mail: \textcolor{blue}{orchidp@iiitd.ac.in}
}
}

\maketitle

\begingroup
  \renewcommand{\thefootnote}{\fnsymbol{footnote}}
  \setcounter{footnote}{0}
  \footnotetext{* Contributed equally as first authors.}
\endgroup

\thispagestyle{firststyle}
\pagestyle{fancy}

\begin{abstract}
This paper investigates the polyglot (multilingual) speech foundation models (SFMs) for Crowd Emotion Recognition (CER). We hypothesize that polyglot SFMs, pre-trained on diverse languages, accents, and speech patterns, are particularly adept at navigating the noisy and complex acoustic environments characteristic of crowd settings, thereby offering a significant advantage for CER. To substantiate this, we perform a comprehensive analysis, comparing polyglot, monolingual, and speaker recognition SFMs through extensive experiments on a benchmark CER dataset across varying audio durations (1 sec, 500 ms, and 250 ms). The results consistently demonstrate the superiority of polyglot SFMs, outperforming their counterparts across all audio lengths and excelling even with extremely short-duration inputs. These findings pave the way for adaptation of SFMs in setting up new benchmarks for CER.
\end{abstract}

\section{Introduction}
Crowd Emotion Recognition (CER) involves the intricate task of predicting collective emotional states in large groups, where emotions are conveyed vocally through cheering, booing, and clapping, and visually through gestures such as waving and synchronized movements. These emotional expressions are prevalent at events like sports matches, concerts, political rallies, and social movements, where they profoundly shape the overall atmosphere and influence both participants and performers. Accurately capturing and interpreting these emotions 
is crucial across fields such as public safety, event management, and social research. Research in CER has predominantly centered around visual modalities, such as facial expressions and body movements~\cite{baig2014crowd, quach2022non, wang2022self, zhang2022semi, zhu2023towards, wu2022crowd}. However, audio-based approaches remain relatively unexamined, highlighting a significant gap in the field. \par

CER through audio poses distinct challenges, largely due to the noisy, spontaneous, and overlapping nature of crowd expressions, particularly in audio streams. Crowds express emotions collectively, but the complexity and dynamism of these environments have left CER underexplored compared to speech emotion recognition (SER), that focuses on recognizing a single individual emotions. Additionally, the limited availability of labeled data also hinders the research into CER through audio cues. Overcoming these obstacles necessitates innovative approaches capable of managing both the linguistic diversity and the noisy, dynamic environments typical of large crowds. To solve this gap, Franzoni et al. \cite{franzoni2020emotional} presented the first audio-based CER dataset. Adding on this, Faisal et al. \cite{faisal2021eslce} used MFCC with Random Foreset classifier and Vision-based foundation models like MobileNetV2 with spectrograms for CER. Anand et al. \cite{anand2024pulse} used CLAP representations with a downstream neural network-based approach for predicting crowd excitement score. In this work, we focus on CER through audio.

Unlike CER, SER has sufficient development particularly due to recent advancements in speech foundation models (SFMs), such as Wav2Vec, HuBERT, pre-trained on large-scale diverse speech datasets, offer promising solutions for performance benefit, data scarcity as well as prevention of training models from scratch \cite{atmaja2022evaluating, chetiaphukan23_interspeech, chen2023exploring}. These models have also excelent in various other speech processing tasks such as speech recognition \cite{arunkumar22b_interspeech}, speaker segmentation \cite{baroudi2024specializing}, shout intensity prediction \cite{fukumori2023investigating}, and so on. Applying such SFMs to CER holds the potential to significantly improve data efficiency and enhance performance in CER, especially within the noisy and diverse environments that characterize real-world crowd scenarios.

To bridge this gap, we investigate the potential of state-of-the-art (SOTA) SFMs for recognizing crowd emotions and we hypothesize that \textit{polyglot (multilingual) SFMs will prove highly effective for CER  by navigating the noisy and complex environments of crowded environments. This can be attributed to their capacity to capture a wide range of pitches, tones, and emotional nuances owing to their pre-training on diverse speech data that encompasses various languages, accents, and speaking styles.} To validate, our hypothesis, we carry out a large-scale comparison of various SOTA SFMs with different downstreams such as SVM, Random Foreset, Fully Connected Network (FCN), and CNN. To the best of our knowledge, we are the first study to explore SFMs for CER.

\begin{figure*}[hbt]
    \centering
    \includegraphics[scale=0.3]{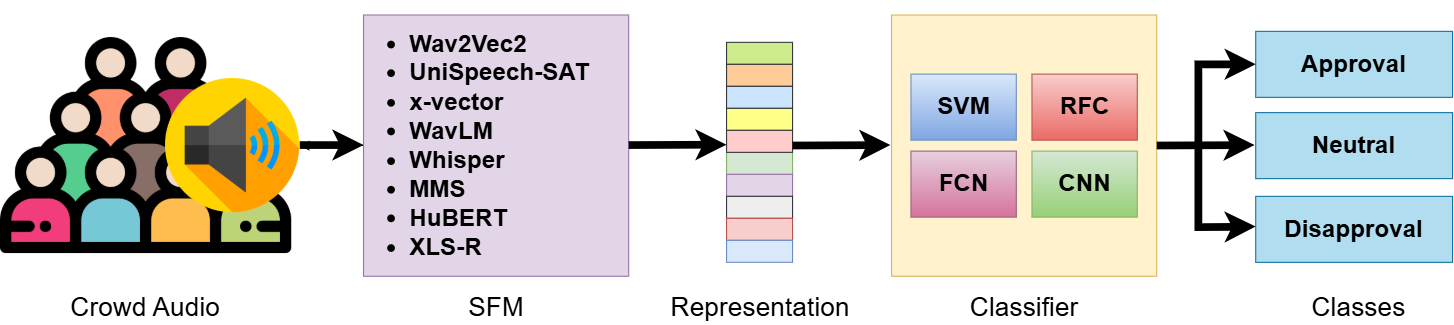}
    \caption{Overview of the proposed system architecture for CER, illustrating the SFM, representation extraction, and classification processes}
    \label{fig:work_flow}
\end{figure*}

\noindent \textbf{To summarize, the major contributions of this study are as follows}:
\begin{itemize}
    \item We conduct comprehensive comparative analysis of SOTA polyglot, monolingual, and speaker recognition SFMs to investigate the efficacy of polyglot SFMs (XLS-R, Whisper, MMS) for CER. Such comparison of SOTA SFMs for CER is the first one to the best of knowledge. Work flow of the paper is given in Figure \ref{fig:work_flow}.
    \item We perform extensive experiments on a benchmark CER corpus and investigate the influence of varying audio durations (1 second, 500 ms, and 250 ms) on CER performance, aiming to identify optimal conditions for emotion recognition and evaluate the potential of SFMs for CER in extremely short duration audios.
    \item Our results reveal that polyglot SFMs consistently outperform their counterparts across all audio lengths, underscoring their superiority for CER tasks, even in extremely short-duration audio segments. 
\end{itemize}


\begin{table}[hbt!]
\centering
\begin{tabular}{c|c|c|c}
    \toprule
    {\textbf{Class}} & {\textbf{\#Clips}} & {\textbf{Duration (s)}} & {\textbf{\#Blocks}} \\ 
    \midrule
    Approval       & 39                                           & 518                                             & 1787                                      \\
    Disapproval    & 15                                           & 118                                             & 388                                       \\
    Neutral        & 15                                           & 1874                                           & 7340                                      \\
    \midrule
    \textbf{Total} & 69                                           & 2510                                           & 9515   \\    
    \bottomrule                              
\end{tabular}
\caption{Dataset statistics: Number of clips per class (\#Clips), total duration (in seconds), and total 1-sec blocks (\#Blocks).}
\label{tab:dataset_statistics}
\end{table}

\begin{figure}[hbt!]
    \centering
    \begin{minipage}[b]{0.1\textwidth}
        \centering
        \includegraphics[width=\textwidth]{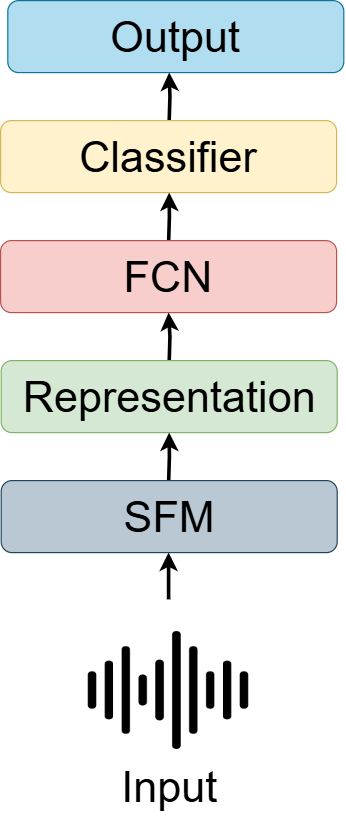}
        \subcaption{FCN}
        \label{fig:fcn}
    \end{minipage}
    \hspace{0.04\textwidth}
    \begin{minipage}[b]{0.1\textwidth}
        \centering
        \includegraphics[width=\textwidth]{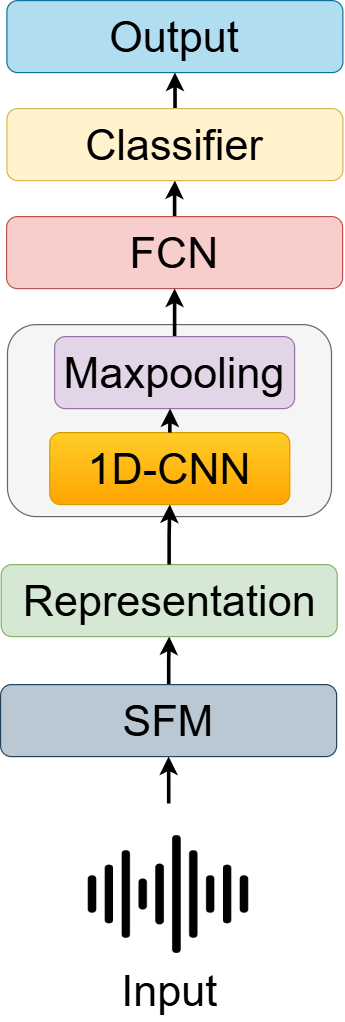}
        \subcaption{CNN}
        \label{fig:cnn}
    \end{minipage}
    \caption{Model Architecture}
    \label{fig:confusion_matrices}
\end{figure}

\begin{table*}[hbt!]
\scriptsize
\centering
\begin{tabular}{l|c|c|c|c|c|c|c|c|c|c|c|c|c|c|c|c} 
\toprule
\textbf{SFM} & \multicolumn{2}{c|}{\textbf{Wav2vec2}} & \multicolumn{2}{c|}{\textbf{UNI}} & \multicolumn{2}{c|}{\textbf{WavLM}} & \multicolumn{2}{c|}{\textbf{HuBERT}} & \multicolumn{2}{c|}{\textbf{x-vector}} & \multicolumn{2}{c|}{\textbf{Whisper}} & \multicolumn{2}{c|}{\textbf{MMS}} & \multicolumn{2}{c}{\textbf{XLS-R}} \\ 
\midrule
 & \textbf{A} & \textbf{F1} & \textbf{A} & \textbf{F1} & \textbf{A} & \textbf{F1} & \textbf{A} & \textbf{F1} & \textbf{A} & \textbf{F1} & \textbf{A} & \textbf{F1} & \textbf{A} & \textbf{F1} & \textbf{A} & \textbf{F1} \\ 
\midrule
\multicolumn{17}{c}{\textbf{1 sec}} \\ 
\midrule
\textbf{SVM} & \cellcolor{green!30}96.94 & \cellcolor{green!30}90.14 & \cellcolor{green!20}95.34 & \cellcolor{green!15}84.83 & \cellcolor{green!10}93.08 & \cellcolor{green!10}73.28 & \cellcolor{green!40}97.87 & \cellcolor{green!35}92.03 & \cellcolor{green!20}95.80 & \cellcolor{green!15}84.55 & \cellcolor{green!50}98.61 & \cellcolor{green!50}95.88 & \cellcolor{green!55}98.75 & \cellcolor{green!50}95.59 & \cellcolor{green!45}98.35 & \cellcolor{green!40}94.04 \\ 
\textbf{RFC} & \cellcolor{green!15}95.35 & \cellcolor{green!10}79.80 & \cellcolor{green!15}95.24 & \cellcolor{green!10}80.67 & \cellcolor{green!10}93.38 & \cellcolor{green!10}76.03 & \cellcolor{green!15}94.07 & \cellcolor{green!10}77.55 & \cellcolor{green!15}95.63 & \cellcolor{green!10}85.24 & \cellcolor{green!40}98.71 & \cellcolor{green!40}94.96 & \cellcolor{green!35}97.45 & \cellcolor{green!30}92.51 & \cellcolor{green!20}96.48 & \cellcolor{green!15}88.81 \\ 
\textbf{FCN} & \cellcolor{green!40}97.29 & \cellcolor{green!35}91.62 & \cellcolor{green!30}96.00 & \cellcolor{green!20}85.75 & \cellcolor{green!15}94.19 & \cellcolor{green!10}78.60 & \cellcolor{green!40}97.87 & \cellcolor{green!35}92.03 & \cellcolor{green!30}97.64 & \cellcolor{green!30}92.80 & \cellcolor{green!50}98.93 & \cellcolor{green!50}96.10 & \cellcolor{green!60}99.06 & \cellcolor{green!55}96.81 & \cellcolor{green!40}98.35 & \cellcolor{green!35}94.04 \\ 
\textbf{CNN} & \cellcolor{green!60}97.51 & \cellcolor{green!55}92.73 & \cellcolor{green!40}96.40 & \cellcolor{green!30}87.17 & \cellcolor{green!25}95.19 & \cellcolor{green!20}84.32 & \cellcolor{green!50}98.09 & \cellcolor{green!45}93.29 & \cellcolor{green!50}98.23 & \cellcolor{green!50}94.61 & \cellcolor{green!55}98.97 & \cellcolor{green!55}96.64 & \cellcolor{green!80}\textbf{99.11} & \cellcolor{green!80}\textbf{96.85} & \cellcolor{green!60}98.71 & \cellcolor{green!55}95.39 \\ 
\midrule
\multicolumn{17}{c}{\textbf{500 ms}} \\ 
\midrule
\textbf{SVM} & \cellcolor{green!30}96.93 & \cellcolor{green!25}89.97 & \cellcolor{green!25}95.69 & \cellcolor{green!20}85.71 & \cellcolor{green!10}92.67 & \cellcolor{green!5}66.04 & \cellcolor{green!40}97.28 & \cellcolor{green!35}90.26 & \cellcolor{green!25}95.56 & \cellcolor{green!20}86.69 & \cellcolor{green!50}98.99 & \cellcolor{green!45}96.47 & \cellcolor{green!55}99.00 & \cellcolor{green!50}97.60 & \cellcolor{green!45}98.41 & \cellcolor{green!40}94.85 \\ 
\textbf{RFC} & \cellcolor{green!25}96.32 & \cellcolor{green!20}86.32 & \cellcolor{green!25}95.71 & \cellcolor{green!20}82.19 & \cellcolor{green!10}92.85 & \cellcolor{green!10}71.57 & \cellcolor{green!15}94.17 & \cellcolor{green!10}75.63 & \cellcolor{green!15}94.57 & \cellcolor{green!10}76.80 & \cellcolor{green!40}97.69 & \cellcolor{green!35}92.49 & \cellcolor{green!35}96.99 & \cellcolor{green!30}90.78 & \cellcolor{green!25}97.00 & \cellcolor{green!20}90.13 \\ 
\textbf{FCN} & \cellcolor{green!35}97.17 & \cellcolor{green!30}90.02 & \cellcolor{green!55}98.61 & \cellcolor{green!40}94.43 & \cellcolor{green!20}93.45 & \cellcolor{green!15}77.08 & \cellcolor{green!50}98.07 & \cellcolor{green!45}92.52 & \cellcolor{green!35}96.48 & \cellcolor{green!25}88.17 & \cellcolor{green!50}98.99 & \cellcolor{green!45}95.81 & \cellcolor{green!55}99.10 & \cellcolor{green!50}96.61 & \cellcolor{green!35}98.68 & \cellcolor{green!30}94.94 \\ 
\textbf{CNN} & \cellcolor{green!40}97.44 & \cellcolor{green!35}91.67 & \cellcolor{green!25}95.52 & \cellcolor{green!20}84.87 & \cellcolor{green!20}94.22 & \cellcolor{green!15}81.00 & \cellcolor{green!50}98.23 & \cellcolor{green!45}94.71 & \cellcolor{green!50}97.58 & \cellcolor{green!45}91.67 & \cellcolor{green!55}98.90 & \cellcolor{green!50}96.17 & \cellcolor{green!80}\textbf{99.19} & \cellcolor{green!80}\textbf{96.66} & \cellcolor{green!50}98.90 & \cellcolor{green!45}95.63 \\ 
\midrule
\multicolumn{17}{c}{\textbf{250 ms}} \\ 
\midrule
\textbf{SVM} & \cellcolor{green!20}95.38 & \cellcolor{green!15}82.83 & \cellcolor{green!20}94.96 & \cellcolor{green!15}83.01 & \cellcolor{green!10}92.48 & \cellcolor{green!5}58.18 & \cellcolor{green!40}97.57 & \cellcolor{green!35}91.86 & \cellcolor{green!20}93.62 & \cellcolor{green!15}81.57 & \cellcolor{green!50}98.81 & \cellcolor{green!45}95.61 & \cellcolor{green!55}98.84 & \cellcolor{green!50}96.34 & \cellcolor{green!45}97.91 & \cellcolor{green!40}92.92 \\ 
\textbf{RFC} & \cellcolor{green!10}92.93 & \cellcolor{green!5}72.87 & \cellcolor{green!10}92.22 & \cellcolor{green!5}70.74 & \cellcolor{green!5}91.16 & \cellcolor{green!5}64.46 & \cellcolor{green!15}95.24 & \cellcolor{green!10}80.68 & \cellcolor{green!15}93.44 & \cellcolor{green!10}72.37 & \cellcolor{green!40}97.98 & \cellcolor{green!35}93.65 & \cellcolor{green!35}97.57 & \cellcolor{green!30}91.86 & \cellcolor{green!20}98.20 & \cellcolor{green!15}93.55 \\ 
\textbf{FCN} & \cellcolor{green!20}95.66 & \cellcolor{green!15}84.86 & \cellcolor{green!20}94.83 & \cellcolor{green!15}81.77 & \cellcolor{green!10}93.25 & \cellcolor{green!5}74.70 & \cellcolor{green!40}97.69 & \cellcolor{green!35}92.05 & \cellcolor{green!20}94.08 & \cellcolor{green!15}79.66 & \cellcolor{green!50}98.95 & \cellcolor{green!45}96.21 & \cellcolor{green!55}98.87 & \cellcolor{green!50}96.47 & \cellcolor{green!45}98.44 & \cellcolor{green!40}94.63 \\ 
\textbf{CNN} & \cellcolor{green!25}96.20 & \cellcolor{green!20}87.25 & \cellcolor{green!25}95.20 & \cellcolor{green!20}83.14 & \cellcolor{green!15}94.08 & \cellcolor{green!10}79.66 & \cellcolor{green!40}97.88 & \cellcolor{green!35}92.38 & \cellcolor{green!20}95.02 & \cellcolor{green!15}83.17 & \cellcolor{green!50}98.97 & \cellcolor{green!45}96.58 & \cellcolor{green!80}\textbf{99.20} & \cellcolor{green!80}\textbf{96.65} & \cellcolor{green!45}98.44 & \cellcolor{green!40}94.63 \\ 
\bottomrule
\end{tabular}
\caption{Evaluation scores of models trained on different SFMs representations; A and F1 stand for accuracy and macro F1-score, respectively. All scores are averaged over five folds and presented as \%; UNI stands for Unispeech-SAT. Light green opacity represents performance levels.}
\label{tableacc}
\end{table*}

\section{Speech Foundation Models}
We leverage SOTA SFMs that excels across various tasks in speech processing across different benchmarks. Each SFM with their distinct technical strengths, provides critical capabilities aligned with our investigation into CER. By integrating these advanced SFMs, our approach harnesses their scalability, robustness, and versatility, making them invaluable for effective CER. Detailed explanation of the Detailed explanation regarding the SFM considered in our study are given below: 

\noindent \textbf{WavLM}\footnote{\url{https://huggingface.co/microsoft/wavlm-base}} \cite{chen2022wavlm} It is a SOTA SFM that outperforms all the other SFMs in various speech procesing tasks in SUPERB benchmark. It learns both speech denoising and masked prediction during training and trained on 960 hours of english librispeech data. We use the base version of 94.70 million parameters.

\noindent \textbf{UniSpeech-SAT}\footnote{\url{https://huggingface.co/microsoft/unispeech-sat-base}} \cite{9747077} employs a contrastive objective alongside multitask learning. Its pre-training follows a speaker-aware approach and is conducted using 960 hours of Librispeech English speech data. We utilize the base variant, which comprises 94.68 million parameters.

\noindent \textbf{Wav2vec2}\footnote{\url{https://huggingface.co/facebook/wav2vec2-base}} \cite{baevski2020wav2vec} employs a self-supervised learning approach, transforming raw audio into latent speech representations through a convolutional feature encoder and Transformer blocks. We consider the base version of 95.04 million parameters trained on 960 hours of Librispeech data in english.

\noindent \textbf{HuBERT}\footnote{\url{https://huggingface.co/facebook/hubert-base-ls960}} \cite{hsu2021hubert} employs a BERT-like masked prediction framework to learn both acoustic and linguistic features effectively. It shows SOTA performance on multiple speech recognition benchmarks in comparison to previous wav2vec2. We use the base version of 94.68 million parameters trained on librispeech 960 hour english data.

\noindent \textbf{XLS-R}\footnote{\url{https://huggingface.co/facebook/wav2vec2-xls-r-300m}} \cite{babu22_interspeech} is a multi-lingual representation learning model based on wav2vec2 architecture and trained on 128 languages. The training datasets comprises of BABEL, Voxlingua107, Commonvoice, MLS, and VoxPopuli. We use the 300 million parameters variant in our work.

\noindent \textbf{Whisper}\footnote{\url{https://huggingface.co/openai/whisper-base}} \cite{radford2023robust} is a multi-task learning SFM based on vanilla transformer encoder-decoder architecture. Pre-training is carried out in 680k hours of multilingual data in 96 languages and in a weakly supervised manner. Whisper shows improve performance than XLS-R for multilingual speech recognition. We utilize the base version of 74 millions parameters. 

\noindent \textbf{Massively Multilingual Speech (MMS)}\footnote{\url{https://huggingface.co/facebook/mms-1b}} \cite{pratap2024scaling} is based on the wav2vec2 architecture and pre-trained on apporximately 1400 languages. It uses around 500k hours of data for its pre-training including FLEURS, MMS-lab, BABEL and solves constrastive learning objective. We use the openly available 1 billion parameters variant.  

\noindent \textbf{x-vector}~\cite{snyder2018x} is specifically designed for speaker recognition and it is a time-delay neural network. However, x-vector has shown its effectiveness in related applications such as SER \cite{chetiaphukan23_interspeech}, shout intensity prediction \cite{fukumori2023investigating}, so we thought it might be helpful for CER and so, we include it in our experiments. It consists of 4.2 million parameters.

\noindent We use frozen SFMs as we want to understand their implicit capacity of understanding CER. We extract representations from the last hidden state through the use of mean pooling with dimensional size of 768 for WavlM, Unispeech-SAT, wav2vec2, HuBERT and 1280 for XLS-R, MMS. For Whisper and x-vector, the dimensions are 512, however, for Whisper, we extract it from the last hidden state of the encoder and discard the decoder. 


\section{Experiments}

\subsection{Benchmark Dataset}

We utilize the only openly accesible dataset for CER, to the best of our knowledge given by Franzoni et al. \cite{franzoni2020emotional} in our study. It was meticulously curated to capture the rich emotional expressions of crowds during high-attendance events, such as sports matches, concerts, political rallies, and public gatherings. 
We segment high-quality audio clips into 1-second blocks with a 0.25 second overlap, yielding a total of 9515 blocks from 69 original clips following Franzoni et al. \cite{franzoni2020emotional}. Each block was assigned  with emotional labels assigned to three distinct categories: \textit{Approval} (cheering, clapping), \textit{Disapproval} (booing, hissing), and \textit{Neutral} (background chatter) as the original audio-clip label. 
More detailed statistics are presented in Table~\ref{tab:dataset_statistics}. Further, we segment these 1-sec blocks into 500 milliseconds (ms), and 250 ms. However, before splitting into ms duration audios, we remove the silence as otherwise some silent audio might be present. We segment into short segments to understand the capability of SFMs for short-segment CER. All the audios were resampled to 16 KHz before passing it through the SFMs. 

\subsection{Downstream Modeling}

We include both classical ML and DL networks as downstream networks with the SFMs. This includes SVM, Random Forest Classifier (RFC), Fully Connected Network (FCN), and CNN. For SVM and RFC, we kept the default hyperparameters. For CNN, it consists of three 1D-CNN layers: the first with 32 filters, the second with 64 filters, and the third with 128 filters, each with a filter size of 3x3, ReLU activation, and a stride of 1. After each convolutional layer, batch normalization and max-pooling with a pool size of 2x2 were applied. The output was flattened and passed through two fully connected layers, the first with 512 neurons and the second with 128 neurons, both using ReLU activation. A dropout layer with a rate of 0.5 was added between the layers to prevent overfitting. The final output layer used softmax activation with 3 neurons for classification. Similarly, the FCN model followed a similar structure, consisting of fully connected layers with 512, 256, and 128 neurons, each using ReLU activation, and a dropout rate of 0.5 was applied to prevent overfitting. The output layer in the FCN model also had 3 neurons with softmax activation. The design of our methodology is depicted in Figure~\ref{fig:work_flow}. Detailed architectural details of FCN and CNN are given in Figure \ref{fig:fcn} and \ref{fig:cnn}. FCN models trainable parameters are from 1 to 3 millions followed by the CNN models with 2.5 to 4 millions.

\subsection{Training Details}
We use Adam as the optimizer and learning rate of 0.001. We employ categorical cross-entropy as loss. Training was conducted over 50 epochs with a batch size of 32. We use 5-fold cross-validation for training and testing. Here, 4 folds are used for training and one fold for testing. Additionally, we make use of early stopping for preventing overfitting. 

\begin{figure*}[!bt]
    \centering
    \begin{minipage}[b]{0.24\textwidth}
        \centering
        \includegraphics[width=\textwidth]{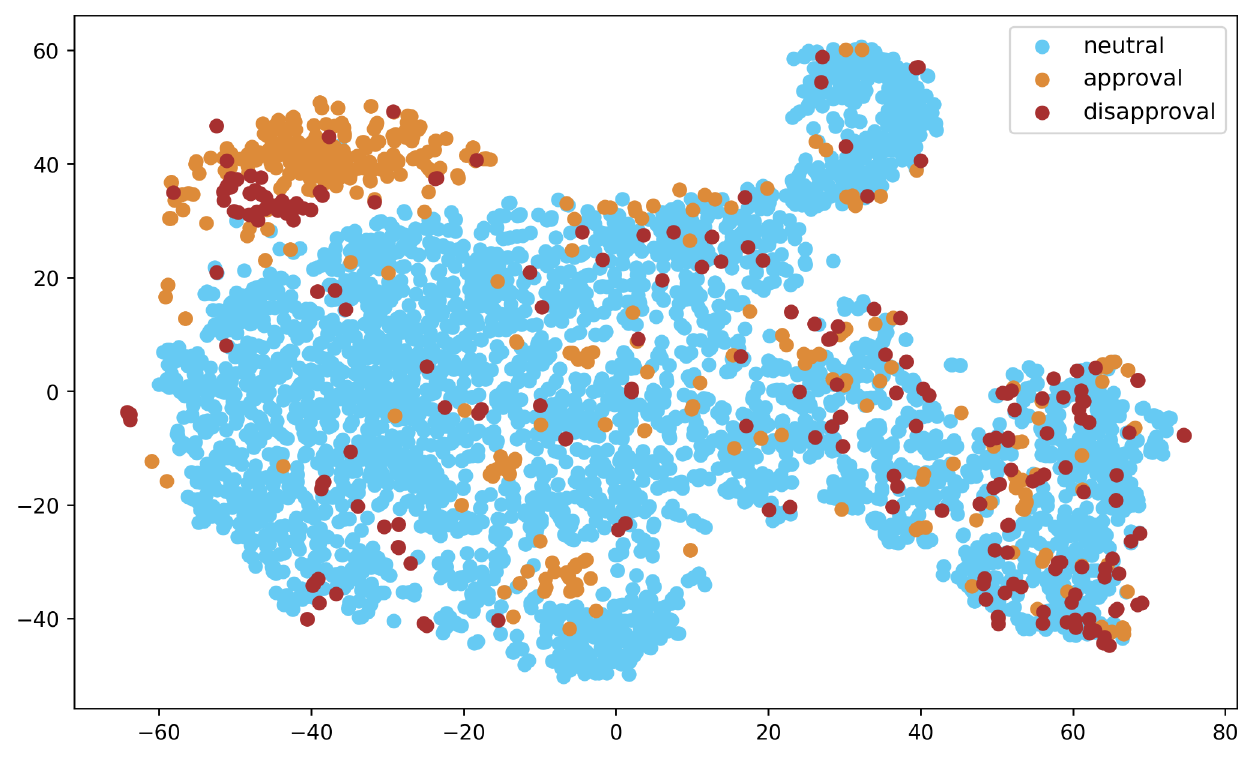}
        \subcaption{}
        \label{fig:fig2}
    \end{minipage}
    \hfill  
    \begin{minipage}[b]{0.24\textwidth}
        \centering
        \includegraphics[width=\textwidth]{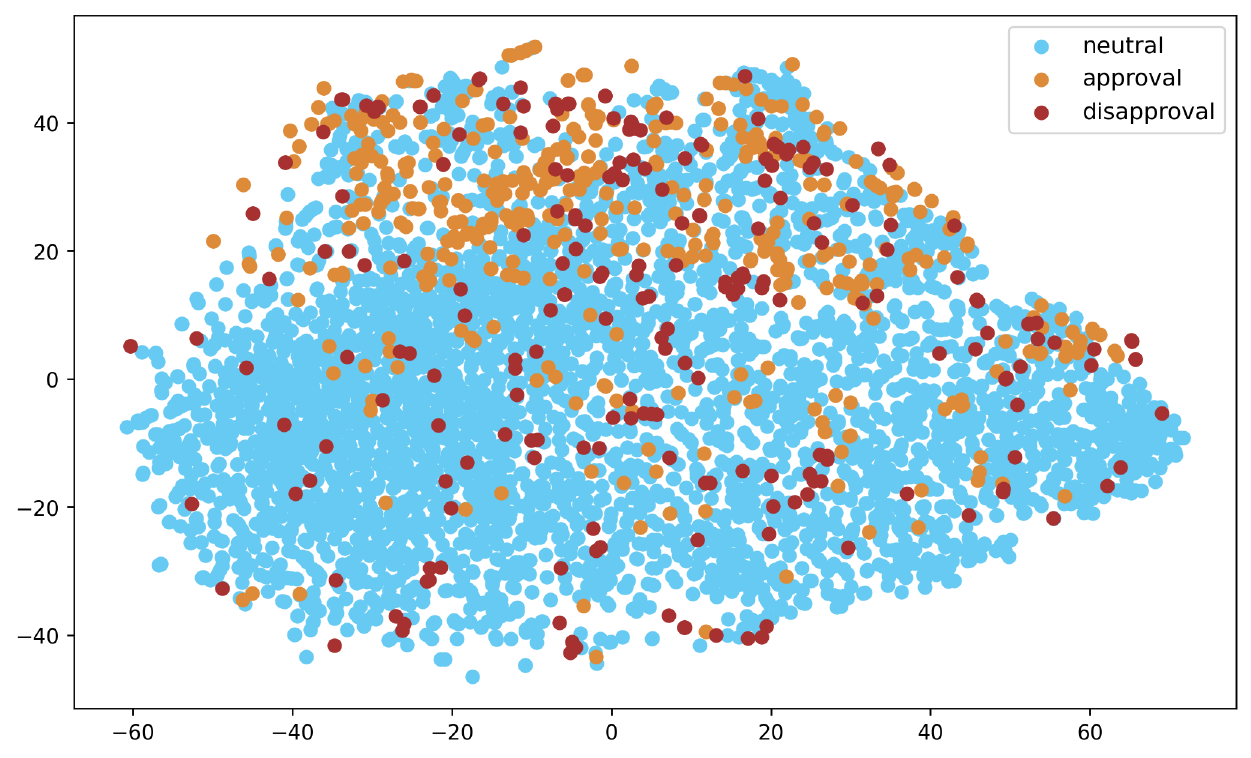}
        \subcaption{}
        \label{fig:fig3}
    \end{minipage}
    \hfill
    \begin{minipage}[b]{0.24\textwidth}
        \centering
        \includegraphics[width=\textwidth]{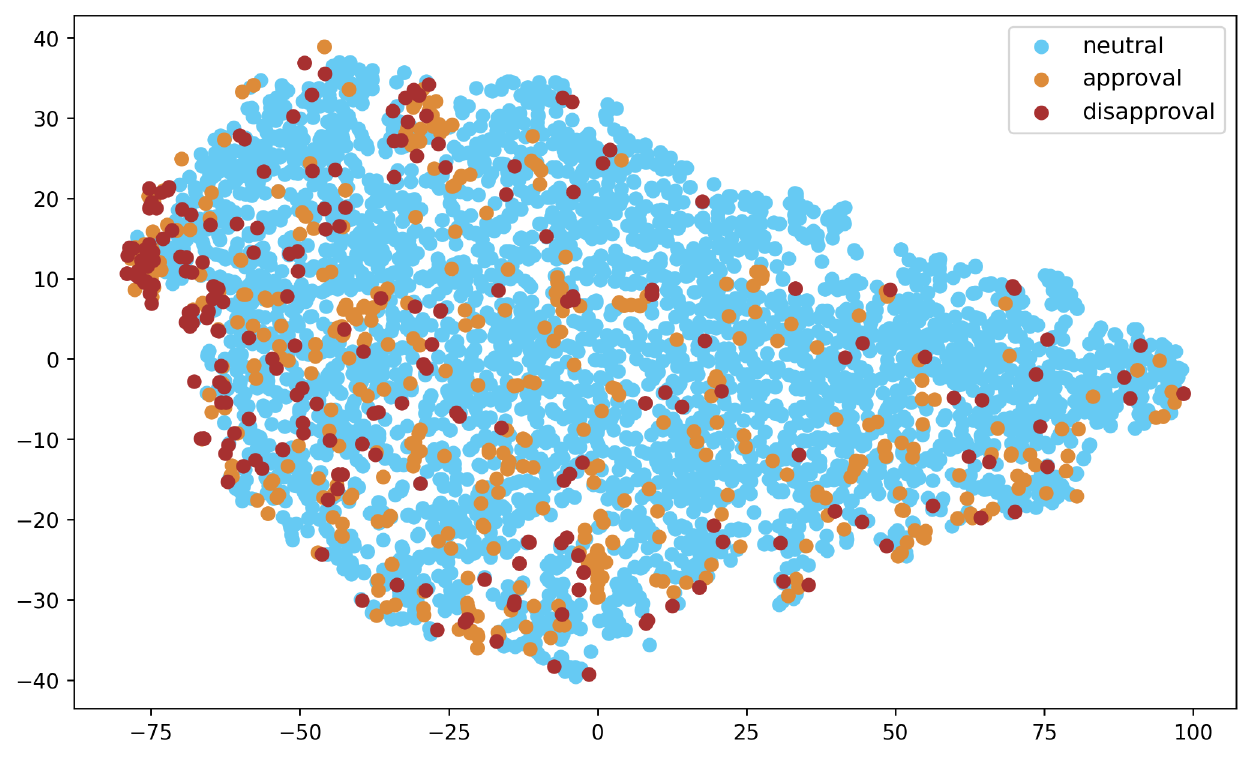}
        \subcaption{}
        \label{fig:fig1}
    \end{minipage}
    \hfill
    \begin{minipage}[b]{0.24\textwidth}
        \centering
        \includegraphics[width=\textwidth]{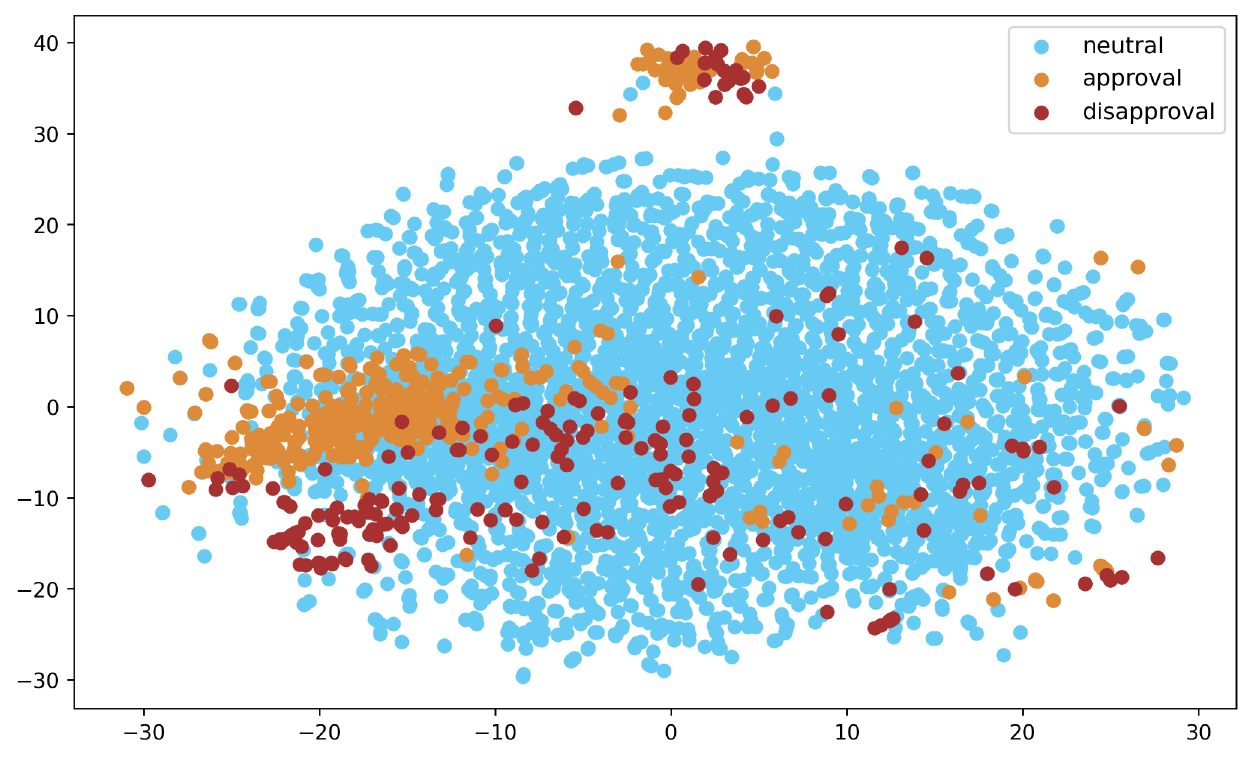}
        \subcaption{}
        \label{fig:fig4}
    \end{minipage}
    
    \vspace{0.4cm}
    
    \begin{minipage}[b]{0.24\textwidth}
        \centering
        \includegraphics[width=\textwidth]{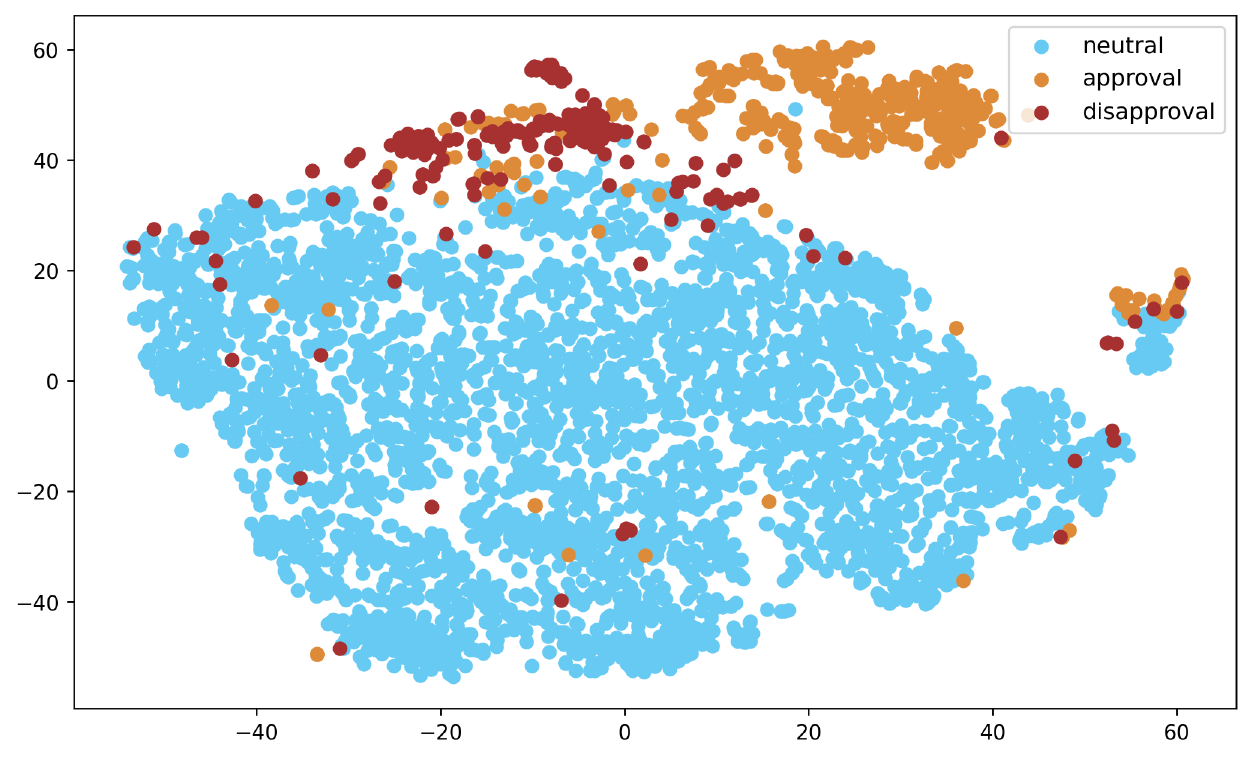}
        \subcaption{}
        \label{fig:fig5}
    \end{minipage}
    \hfill  
    \begin{minipage}[b]{0.24\textwidth}
        \centering
        \includegraphics[width=\textwidth]{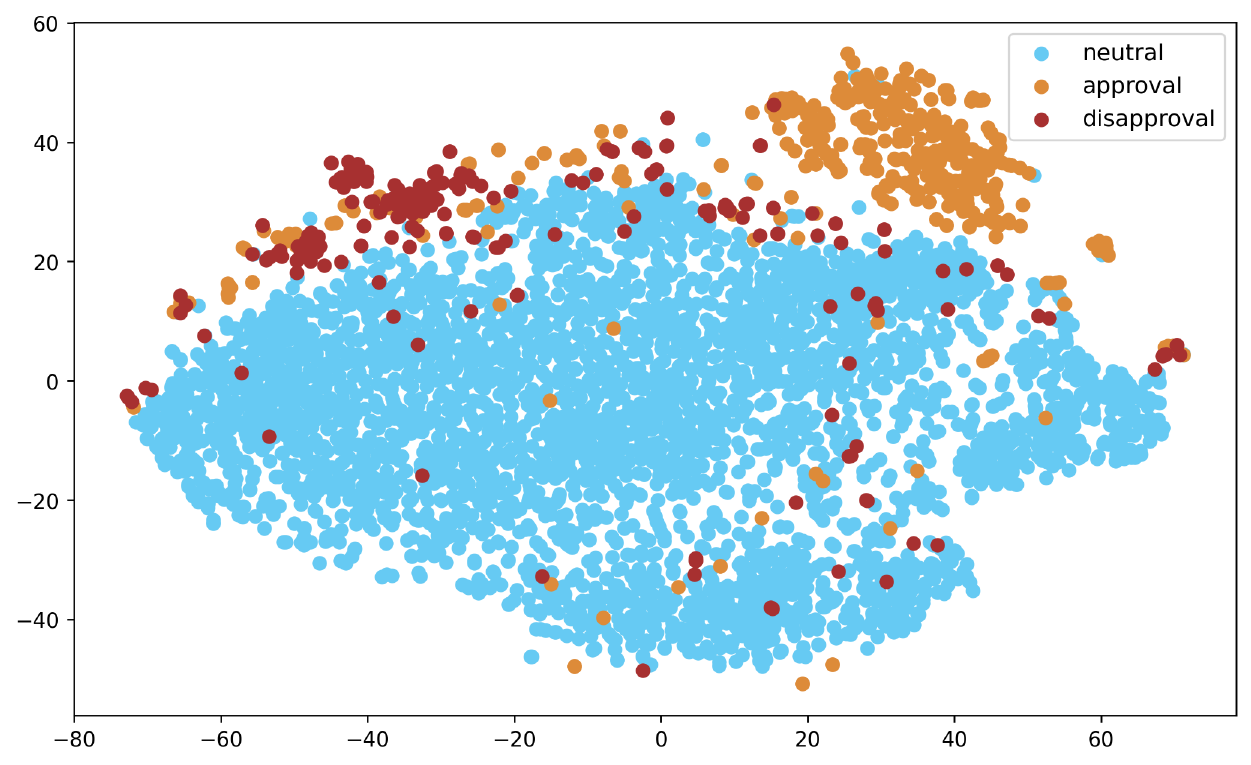}
        \subcaption{}
        \label{fig:fig6}
    \end{minipage}
    \hfill
    \begin{minipage}[b]{0.24\textwidth}
        \centering
        \includegraphics[width=\textwidth]{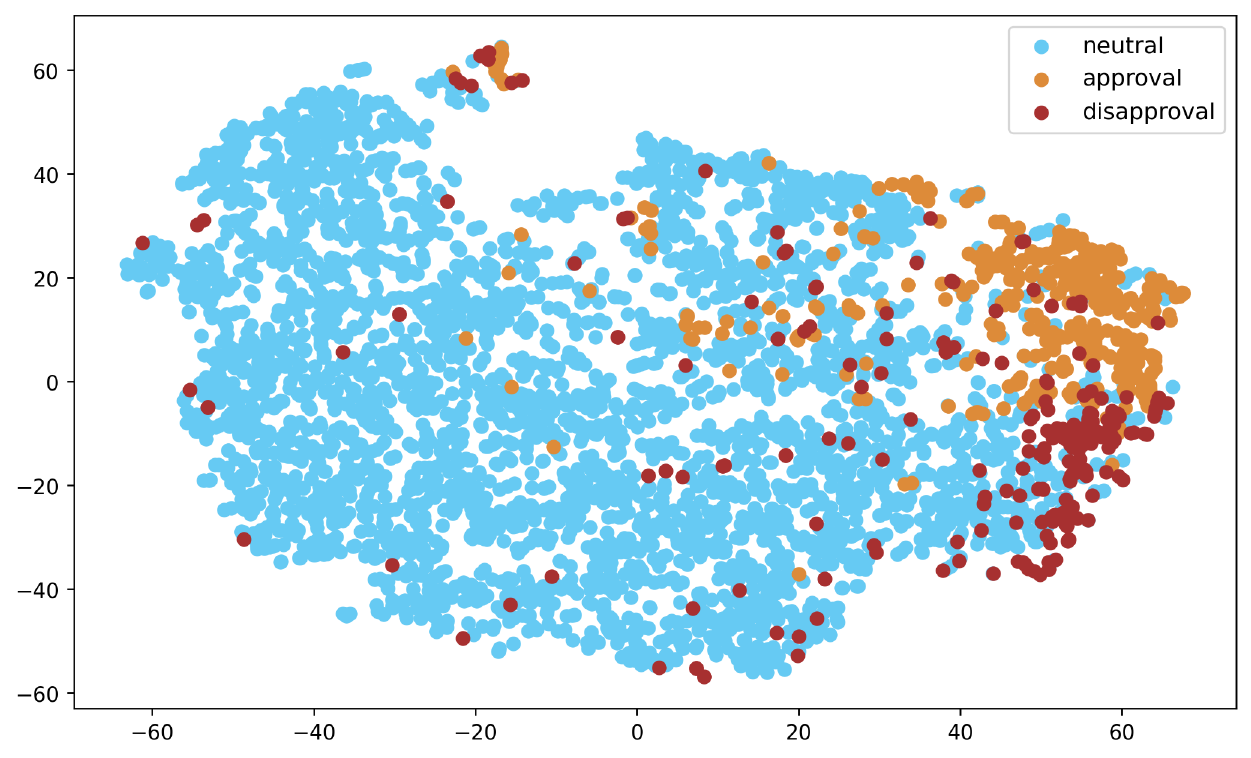}
        \subcaption{}
        \label{fig:fig7}
    \end{minipage}
    \hfill
    \begin{minipage}[b]{0.24\textwidth}
        \centering
        \includegraphics[width=\textwidth]{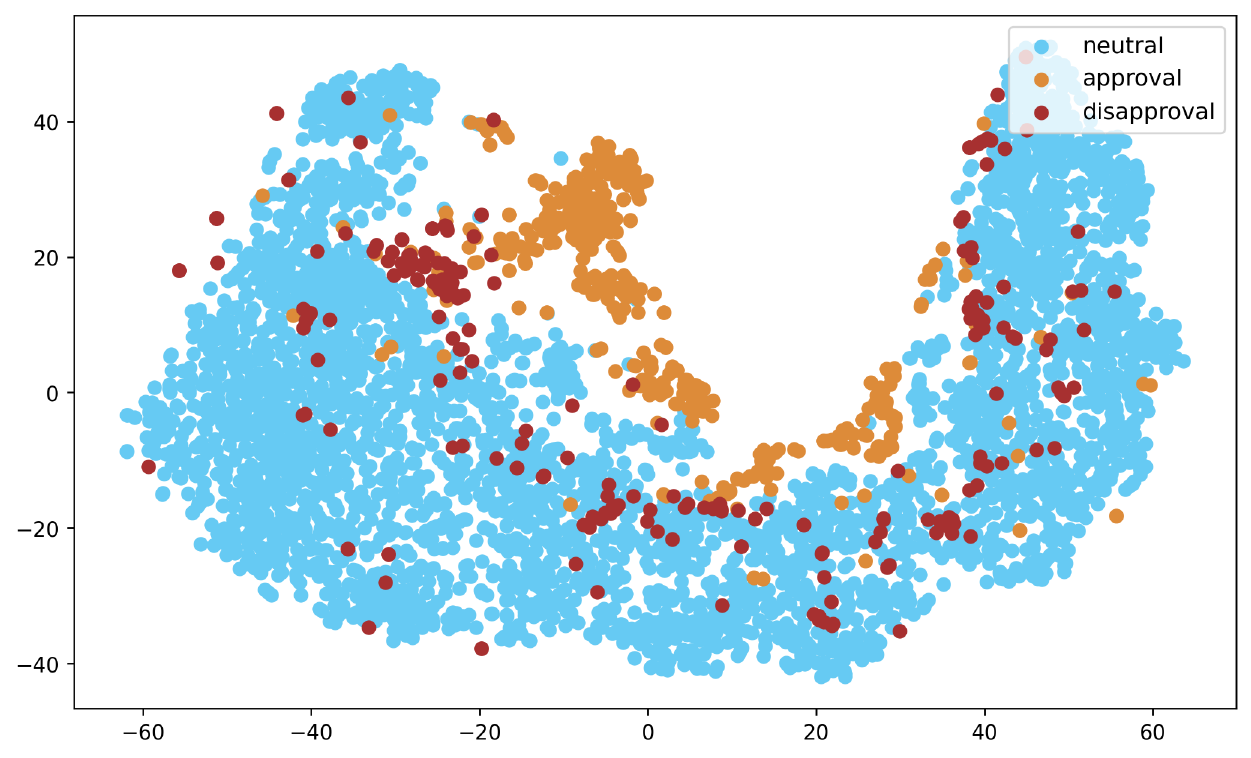}
        \subcaption{}
        \label{fig:fig8}
    \end{minipage}

    \caption{t-SNE plot visualization of different SFMs: (a) Wav2vec2, (b) Unispeech-SAT, (c) WavLM, (d) x-vector, (e) Whisper, (f) MMS, (g) HuBERT, and (h) XLS-R}
    \label{fig:tsne}
\end{figure*}

\begin{figure}[!bt]
    \centering
    \begin{minipage}[b]{0.22\textwidth}
        \centering
        \includegraphics[width=\textwidth]{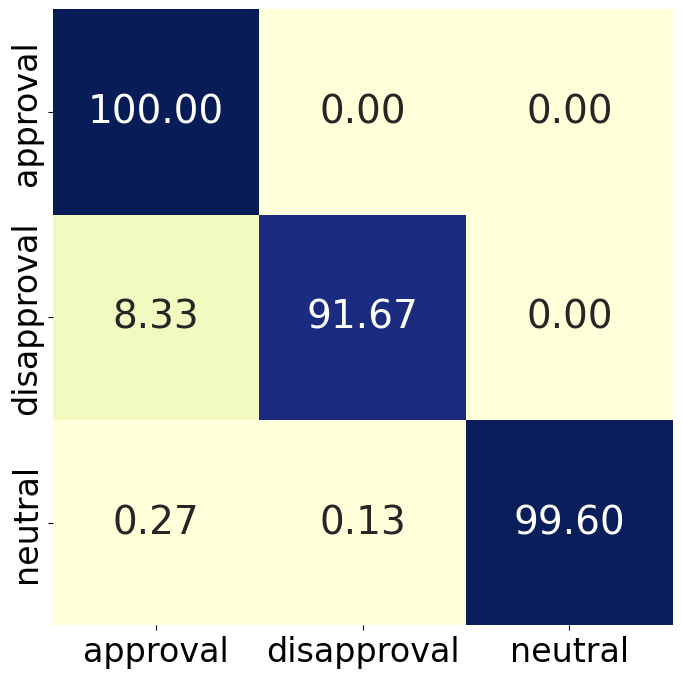}
        \subcaption{}
        \label{fig:fig1}
    \end{minipage}
    \hspace{0.01\textwidth}
    \begin{minipage}[b]{0.22\textwidth}
        \centering
        \includegraphics[width=\textwidth]{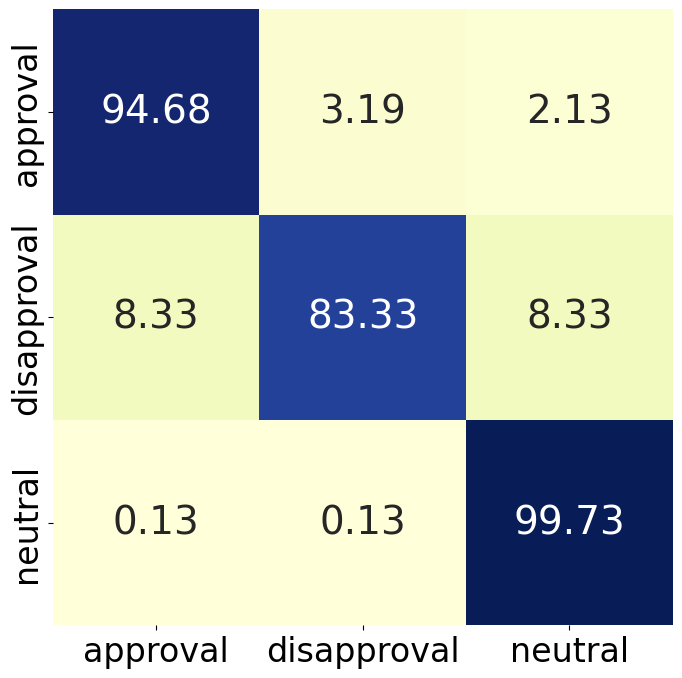}
        \subcaption{}
        \label{fig:fig2}
    \end{minipage}
    \caption{Confusion matrices for 1 sec duration: (a) MMS (b) WavLM; The y-axis represents True Values, while the x-axis denotes Predicted Values.}
    \label{fig:confusion_matrices}
\end{figure}

\begin{figure}[!bt]
    \centering
    \includegraphics[width=0.9\linewidth]{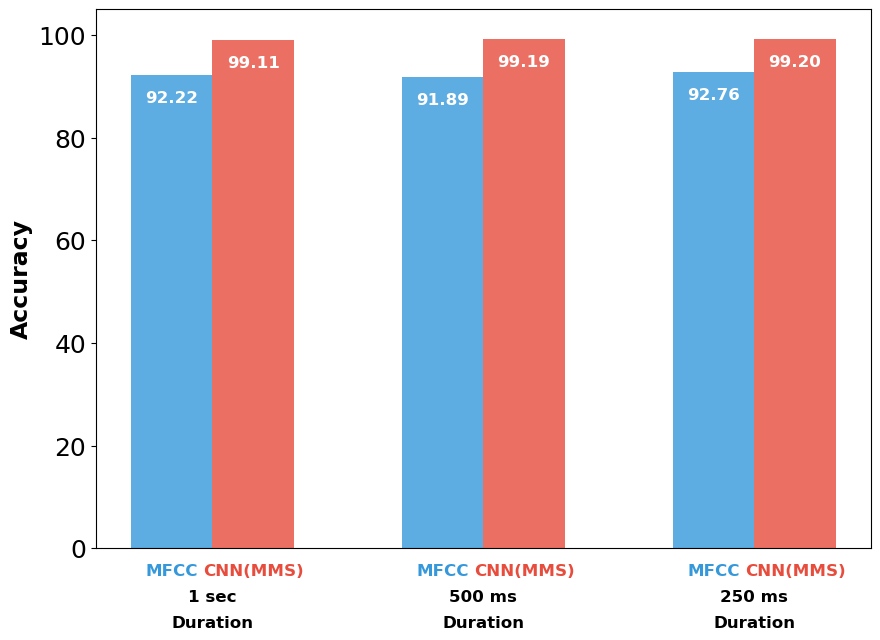}
    \caption{Comparison of the best models with baseline MFCC; CNN (MMS) represents CNN model trained with MMS representations; Scores presented are in \%}
    \label{mfcc}
\end{figure}

\subsection{Experimental Results}

Table~\ref{tableacc} provides the evaluation scores of different downstream models trained on top of SFMs representations. For all the audio durations (1 sec, 500 ms and 250 ms), we can see that polyglot SFMs (XLS-R, Whisper, MMS) completely dominate other SFMs including monolingual and speaker recognition SFMs for CER. This validates our hypothesis that \textit{polyglot SFMs will excel in CER due to their ability to capture diverse pitches, tones, and emotional variations. This stems from their pre-training on a broad spectrum of speech data, covering multiple languages, accents, and speaking styles.} Overall CNN models shows superior performance in comparison to other downstreams (SVM, RFC, FCN) with different SFMs in all the audio durations. For the 1-sec audio segments, MMS demonstrated the topmost performance, achieving an accuracy of 99.11\% and an F1 score of 96.85\% with CNN. Among the polyglot SFMs also, MMS is top and this can be attributed to its larger size of 1 billion parameters allowing it capture the acoustic characteristics required for CER in a much better manner. For 500 ms and 250 ms also, we observe clear dominance of MMS over other SFMs including polyglot SFMs. MMS reported accuracy of 99.19\%, 99.20\% and F1 score of 96.66\%, 96.65\% for 500 ms and 250 ms respectively with CNN. \par

Among the monolingual SFMs (Wav2vec2, Unispeech-SAT, WavLM, HuBERT), HuBERT showed the best results for all the audio durations. It reported accuracy of 98.09\%, 98.23\%, 97.88\% and F1 score of 93.29\%, 94.71\%, 92.38\% for 1 sec, 500 ms, and 250 ms respectively. This could point towards its ability in capturing the diverse crowd sounds in a superior manner. One interesting observation is the performance of x-vector. Despite being a very small SFM comprising of only 4.2 million paramters, it shows comparative performance in comparison with monolingual PTMs in some instances. This behavior can be traced to its speaker recognition providing ability to capture diverse pitches, tones, and so on speech characteristics for effective CER. We visualize t-SNE plots of the raw representations from the last hidden state of the SFMs in Figure~\ref{fig:tsne}. These plots reveal clearer and more distinct clustering for polyglot SFMs across different emotional categories, amplifying the superior performance observed in the results. We also plot the confusion matrices of MMS and HuBERT in Figure \ref{fig:confusion_matrices} for 1 sec duration. These findings underscore the potential of leveraging polyglot SFMs to significantly enhance the performance of CER systems and establishing a solid foundation for future research. \newline
\noindent \textbf{Comparison with MFCC Baseline:} As MFCC is one of the most used input representation for speech and audio processing and used by research works as a baseline for evaluating their proposed methods performance \cite{rathod23_interspeech, charola23_interspeech}, we also give a comparison of the best models with baseline MFCC features to understand the effectiveness of polyglot SFMs. We keep the modeling and training details same for experiments with MFCC as set for experiments with SFMs representations. The comparison is presented in Figure \ref{mfcc}. The best model CNN(MMS) shows superior performance in comparison to baseline MFCC feature.

\section{Conclusion}
In this study, we investigated the effectiveness of polyglot  SFMs for CER. We hypothesized that polyglot SFMs, pre-trained on diverse languages, accents, and speech patterns, would be particularly adept at handling the noisy and complex acoustic environments of crowds, offering a distinct advantage for CER. Through extensive experiments on a benchmark CER dataset with varying audio durations (1 sec, 500 ms, and 250 ms), our analysis confirmed that polyglot SFMs consistently outperformed monolingual and speaker recognition SFMs, demonstrating superior performance even with short-duration inputs. These findings reinforce the potential of polyglot SFMs in CER and set a foundation for future research. Our study will also act as a guide for selection of SFMs for CER and related applications.

\printbibliography

\end{document}